# Google-based Mode Choice Modeling Approach

Zohreh Ghasemi
University of Illinois at Chicago

## Abstract

Microsimulation based frameworks have become very popular in many research areas including travel demand modeling where activity-based models have been in the center of attention for the past decade. Advanced activity-based models synthesize the entire population of the study region and simulate their activities in a way that they can keep track of agents' resources as well as their spatial location. However, the models that are built for these frameworks do not take into account this information mainly because they do not have them at the modeling stage. This paper tries to describe the importance of this information by analyzing a travel survey and generate the actual alternatives that individuals had when making their trips. With a focus on transit, the study reveals how transit alternatives are limited\unavailable in certain areas which must be taken in to account in our mode choice models. Some statistics regarding available alternatives and the constraints people encounter when making a choice are presented with a comprehensive choice set formation. A mode choice model is then developed based on this approach to represent the importance of such information.

## Introduction

Traditionally transportation related models were designed to deal with aggregate information or specific market segments; however with advancement in technology, much more information has become available to researchers as the demand for more policy sensitive models has increased. These advances have made microsimulation frameworks very popular in the past two decades (1-6).

Although Microsimulation frameworks could potentially be sensitive to many important factors, in which modelers and planners have always been interested, they also introduce many challenges and issues that if not taken into account, they could highly bias their outcomes. In the area of regional travel demand modeling, Activity-based models have been the center of attention for the past decade and many researchers have proposed and developed different frameworks (1, 7, 8). These microsimulation frameworks try to substitute trip-based four-step approaches, which were highly aggregate, by simulating individuals activities throughout the day (or even week or month) while conserving many institutional and resource constraints (1).

In this type of frameworks, census public data is utilized to synthesize the entire population of the study region along with their socio-demographic/economic attributes (8). For most of the currently developed activity-based models, this information is used during the activity generation and scheduling part and is not carried to the traffic assignment step of the program. However, more advanced frameworks take advantage of this information even during the traffic assignment. The state of the art studies even have an integrated traffic assignment module that is run concurrently with the activity generation module. The agents have full access to the network information and also availability of resources at each time step so that they could update their decisions accordingly. This level of details is available to all of the models within the larger framework.

The traditional frameworks, however, were built mostly for a limited market segment and were not necessarily based on disaggregate and accurate information(8). Therefore, enforcing the integrity of resources/constraints was not feasible in such simulation environments; hence related models were designed to address these limitations and information using statistical methods. Other fields have also used similar algorithms (11, 17). For example, four-step travel demand models are designed to investigate the demand for specific periods of a day (mainly rush hours). They treat trips in aggregate numbers so that it is not possible to ensure those who have left their home in the morning would return to their places in the afternoon. It is not also possible to ensure members of a household are not using more vehicles than the number of vehicles they own. On the other hand, activity-based models are designed to investigate activities and travels of individuals for a longer duration (usually more than 24 hours). All individuals could be tracked during this period and institutional and resource constraints could be enforced. Therefore, for example, a mode choice model in an activity-based framework could take advantage of the knowledge of vehicle unavailability at a time, while a mode choice model used in a four-step framework, could just rely on the number of vehicles a household owns.

Moreover, advanced technology has enabled researchers to collect accurate information, in terms of household information, their travel behavior and decision making process and also network conditions as well as other information that could affect individuals' travel and its attributes. The technology, powerful processors, multithreading and distributed computing techniques have also made complicated microsimulation tasks much faster and more feasible. They could take advantage of accurate network and public transit information and take them into account during the simulation. For example in such environments it is feasible to use the location of individuals, the transit system availability and proximity at the time of interest to make an alert decision when choosing a mode for travel.

With this introduction, the purpose of this study is to describe the advantages and importance of accurate data and resource constraints including transit in the models within activity-based frameworks through a case study. In the following sections first a brief discussion of mode choice modeling is presented and then the study area and the data that has been used is described. Next the choice set formation is elaborated followed by model development and result discussion before the conclusions are presented.

## Mode Choice Modeling

Mode choice is a key module for long term travel demand studies (i.e. regional travel demand models), and also for many short term studies (e.g. congestion pricing, toll value analysis). In traditional four-step models, after trips are generated and distributed between zones, mode choice analysis is conducted, before the traffic assignment step, to find the share of each mode for passenger movement (9). In tour-based or activity-based models, the selection of mode for each trip (or tour) is conducted at different steps of the larger framework. For example in many tour-based, and also activity-based models, similar to the four-step ones, the mode is determined once the tour is formed (10), however in some others, first the mode is selected and then the tour is formed. For more advanced travel demand models (i.e. dynamic activity-based models), the order of travel related choices (e.g. mode, destination, departure time, party composition) for each individual could vary from one to another. For example, in the ADAPTS dynamic activity-based travel demand model, for each agent, a model is run to determine the order of choices and then at the specified order each travel attribute is determined (11).

For building a mode choice model a travel survey is usually conducted and then the survey is analyzed to compile information (e.g. Socio-demographic and socio-economic, available resources, trips attributes) needed for developing the model. Meanwhile, other data including (e.g. network characteristics and geographical information) are added to the survey data to investigate their potential impact on trips attributes. Moreover, since the purpose of the model is to select one of the available modes, the information of alternative modes is also required for this process. Since the characteristics of alternative modes (e.g. feasibility, travel time, cost) are mostly unknown, they have to be estimated. Once the data is gathered, a statistical method (e.g. discrete choice modeling) is used to analyze it and develop the model.

Although mode choice is one of the most studied travel attributes, if not the most, and there is much literature on the development process, the literature have mostly focused on adopting different statistical methods while making them more and more complex (15). However, there are many details in generating the input dataset itself which could highly reduce the accuracy or even bias the results if not enough effort is made to prepare them. A well designed survey could capture many of details, events and conditions led to adopting the observed mode. Some other information could also be gathered from transportation related data sources (e.g. transportation network information, population and spatial information). Resource and constitutional constraints, however, are among many factors/information that potentially could have large impact on individuals' decisions but are usually ignored or roughly estimated. Although these constraints play a very important role in selecting a mode, the conceptual design and modeling approach in traditional mode choice methods greatly limits their capability in taking them into account. Some of the constraints could be inferred from the survey, with deeply analyzing the collected data.

Statistical models similar to Discrete Choice Models are designed to capture the random taste or unobserved factors of individuals when choosing an alternative(7,6). However, the more the logical thinking in selecting an alternative is known and imposed on the data, the more the model could become accurate while the random taste becomes actually a random variable. This also lets the factors that indeed differentiate between alternatives, for specific conditions, to stand out and become more significant. Following sections elaborate a case study that has been conducted to present this information.

## Study Area and Data

Public transportation in the Metropolitan of Chicago is made available by three major agencies including CTA (with both Bus and Subway fleet) which serves the city of Chicago residents, Pace system which mainly serves intra and inter Suburban trips, and also the suburb to city demand for travel, and finally Metra, which mainly gives service to suburbs to city trips with its heavy rail fleet of cars. Figure 1 depicts the area where each agency serves.

The data used in this study comes from the CMAP Travel Tracker Survey which is a comprehensive travel and activity survey conducted by Chicago Metropolitan Agency for Planning (CMAP). The survey was designed for use in a regional travel demand model and was conducted for over a 14 months period beginning in January 2007. More than 14,000 households participated in the survey and recorded their activities and travels for one or two days. At the end, more than 218,005 activities were recorded. The survey dataset was analyzed, invalid records were eliminated and trips and home-based tours were formed and linked to socio-demographic information of individuals.

**Figure 1- Chicago Public Transportation Network**

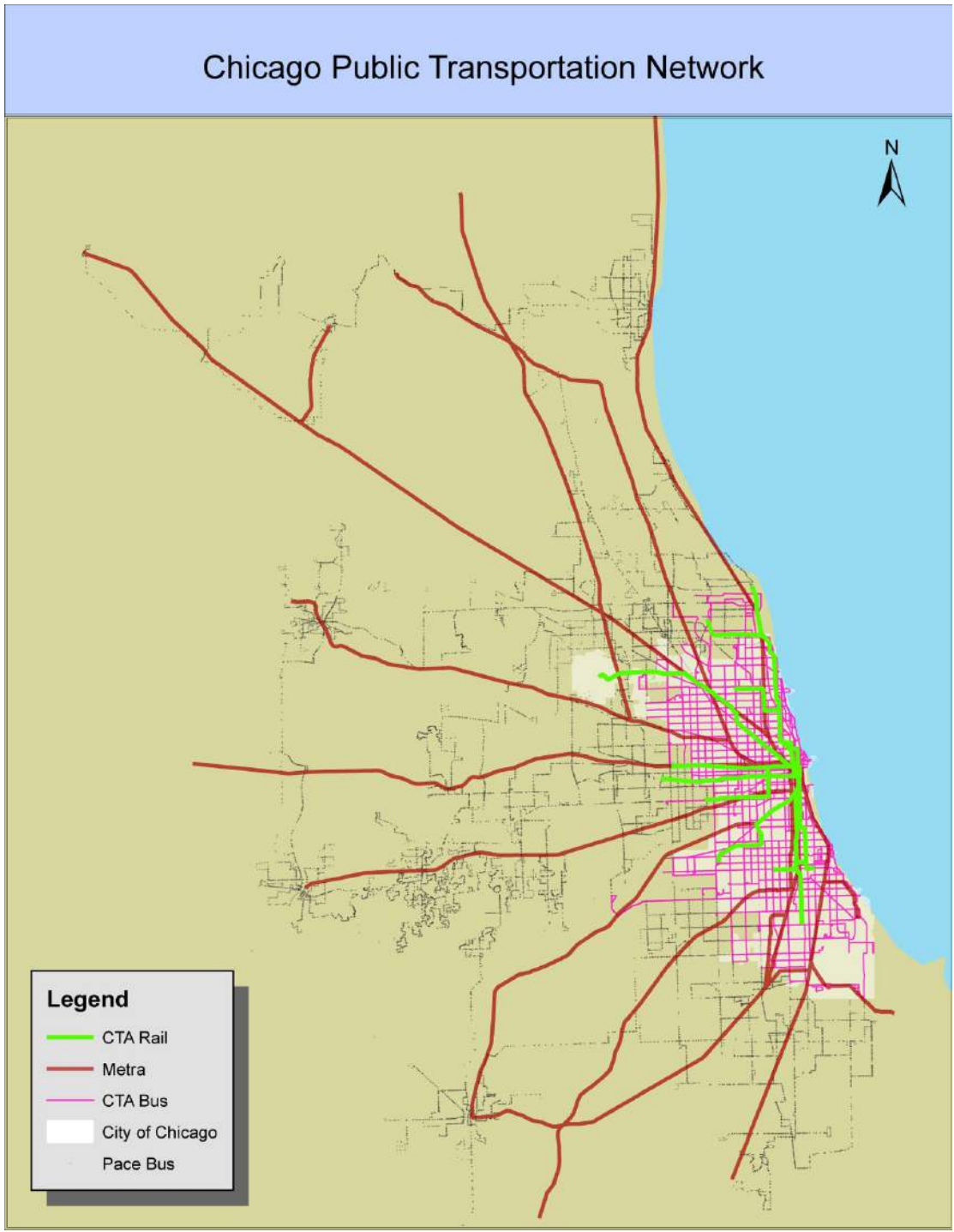

Figure 2 represents the ridership percentage in the greater Chicago area block groups based on the CMAP data. As expected, most of the suburban areas have ridership as low as 5%; however, the city area has ridership as much as 50% or higher in certain block groups. One of the resource constraint parameters in mode choice modeling could be the availability of transit. In case, transit is not offered for a number of trips in the data, the alternative modes must be accommodated based on such constraints.

Figure 2- Transit Ridership in the Greater Chicago Area

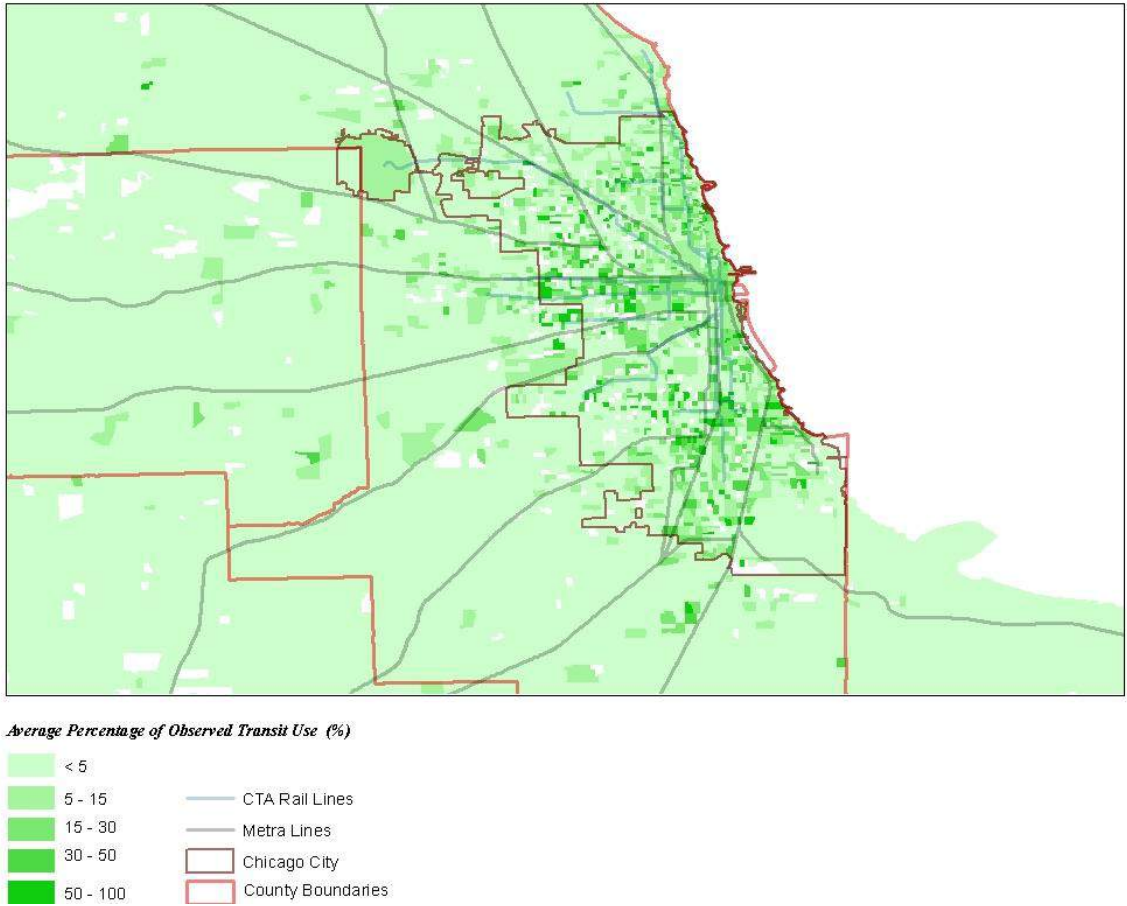

## Choice Set Formation

To develop a mode choice model, first, it was needed to select the alternatives that were going to be used in the model. As it was mentioned before, each transit agency targets specific type of trips and spatial location in the region, their fare structure, quality of ride, distance between stations/stops, and many other factors is consistent within each agency. Therefore, it was decided to make each agency as a separate alternative to travelers in addition to walk, bike, auto drive and auto passenger modes. Since the access distance to Metra stations significantly differs from place to place in the suburbs and many people tend to drive to transit stations, while many others simply walk, it was decided to separate these two from each other.

Before developing a mode choice model for the region, a comprehensive descriptive analysis was conducted to investigate how to proceed with the modeling step. A few factors seemed to be effective on mode selection, and they have been tested in model development. One of the analyses that were conducted was the distribution of modes, when previous mode of a trip was auto-drive (Table 1). In this table, trips are categorized into two groups, "return-home" and "others". As it can be seen, almost all return-home trips and more than 96% of others trips use the same mode as their previous mode (i.e. auto-drive). It is highly possible that if the sub-tours are formed, the rate of auto-drive mode will be higher than 96.5%. Moreover, when an individual has not taken a vehicle in the first leg of a tour, it is rational to assume auto-drive mode is not an option for rest of the tour (there are always exceptions, but exceptions are negligible). These represent a constraint that individuals encounter when choosing a mode and confirm the rational expectation that when a transit or auto-drive tour starts, the mode usually does not change (unless for sub-tours; access mode has been taken into account). This conclusion highly impacts the structure of a mode choice model in terms of adopting tour-based approach instead of trip-based approach or at least taking the previous mode of a trip into account in model development.

Table 1- Distribution of mode when previous trip mode is auto drive

|  | return home | | others | |
| --- | --- | --- | --- | --- |
| **Mode Name** | count | Percentage | count | Percentage |
| **Auto / Van / Truck Driver** | 33,885 | 99.36% | 30,061 | 96.54% |
| **Walk** | 42 | 0.12% | 558 | 1.79% |
| **Auto / Van / Truck Passenger** | 144 | 0.42% | 344 | 1.10% |
| **Metra Train** | 4 | 0.01% | 61 | 0.20% |
| **CTA Train** |  | 0.00% | 33 | 0.11% |
| **School Bus** |  | 0.00% | 23 | 0.07% |
| **Taxi** | 2 | 0.01% | 18 | 0.06% |
| **OTHER** |  | 0.00% | 9 | 0.03% |
| **More than one transit provider** | 3 | 0.01% | 9 | 0.03% |
| **Bike** | 22 | 0.06% | 7 | 0.02% |
| **Private shuttle bus** |  | 0.00% | 6 | 0.02% |
| **CTA Bus** | 1 | 0.00% | 4 | 0.01% |
| **Pace Bus** |  | 0.00% | 4 | 0.01% |
| **Dial a ride/Paratransit** | 1 | 0.00% | 1 | 0.00% |

Another type of constraint that should be taken into account is whether a household member had access to a vehicle for travel. In traditional mode choice models, this has been represented with household number of vehicles (because in simulation this information was not available). Although this question is sometimes asked in surveys, including this survey, but data mining in the dataset and tracking households' members' activities revealed many other cases where the vehicle is being used by another household member when the trip was started (or in proximity of that time). Therefore other family members have to seek other sources of transportation. Although this is very intuitive and a rational assumption, but in most trip based models and models based on datasets that do not carry this information, this critical factor is ignored. Table 2 represents the distribution of mode for households with just one vehicle, when that vehicle is being used by another household member. Incorporating this information in model development (i.e. eliminating the auto-drive alternative) could improve models' accuracy.

It is worth mentioning that according to the data, there are people who drove a vehicle even when the only vehicle of the family was being used. However, this does not necessarily mean that in simulation, the auto-drive mode should be provided as a feasible alternative in mode selection model for such cases. Instead, a more detailed model which could take this type of situations into account could be implemented or a distribution could be used to represent such cases.

Table 2- Selected mode when another household member is using the only vehicle (HHVEH=1)

| Mode | Count | Total | |
|---|---|---|---|
| Walk | 746 | 4,323 | 17% |
| Bike | 64 | 407 | 16% |
| Auto / Van / Truck Driver | 444 | 24,798 | 2% |
| Auto / Van / Truck Passenger | 5,840 | 7251 | 81% |
| CTA Bus | 127 | 688 | 18% |
| CTA Train | 77 | 640 | 12% |
| Pace Bus | 12 | 71 | 17% |
| Metra Train | 72 | 529 | 14% |
| Private shuttle bus | 7 | 63 | 11% |
| Dial a ride/Paratransit | 12 | 19 | 63% |
| School Bus | 163 | 381 | 43% |
| Taxi | 25 | 150 | 17% |
| Local Transit (NIRPC region) | 3 | 16 | 19% |
| More than one transit provider | 31 | 270 | 11% |
| OTHER (SPECIFY) | 3 | 29 | 10% |

Since a discrete choice modeling approach was selected as the modeling method, after the observed trip records were analyzed, the alternative modes were needed to be generated. To generate the alternative modes the authors mainly relied on Google Map website. A software application was developed to take advantage of Google Maps API (https://developers.google.com/maps/documentation) and also RTA's Goroo TripPlanner website (http://tripsweb.rtachicago.com) in order to find the travel times of each trip for Walk, Bike, Transit and Drive modes (alternative modes). In order to find transit travel times, Google relies on the information that each transit agency provides in GTFS format (https://developers.google.com/transit/gtfs). Each transit agency uploads its vehicles schedule, route, and fare structure while Google takes advantage of this data to find the best route. The website also offers travelers to choose one of three transit methods: Bus, Subway, Rail and for this study all three options were selected and the suggested route information were recorded for the alternative modes of the observed trips (16).

Moreover, since all legs of the trips were known (when more than one transit is used), the fare associated with each trip could be exactly calculated, even though each agency has a complicated fare system in terms of transfers, origin and destination station, and age of the travelers. The software was also designed to query the travel times of trips at the same day of week and exact time of the day that original trips were made. For driving mode, this process was repeated a few times, and the average was used. It is worth mentioning that although travel times in 2014 are not necessarily as they were in 2007-2008, when the original trips were made, but this approach seems to be the best available estimation method.

After collecting alternative modes information, a comprehensive analysis was conducted. The first observation noticed was that not all alternatives are available for many trips. One reason for this was that for many cases there is no transit stop or station close to the origin or destination and the other reason is that an agency may not operate around the time of the trip. The conclusion for both cases is that the alternative is not available and should be eliminated from the feasible alternatives. As it can be seen in Table 3 for the case of CTA, almost 80% of the trips did not have any CTA alternative, close to 75% of the trips did not have Pace alternative, 75% did not have a Rail alternative, while 47% of trips did not have any transit alternative. Figure 2 shows the block groups where transit was barely available for the observed trips. The red block groups show that less than 30% of the observed trips had at least a transit alternative at the time when the trips occurred. In other words, the red block groups represent the transit deserts in the Chicago area and could be the target of future transit projects.

Table 3- Transit Alternative Availability for Observed Trips

|  | CTA (Bus and Subway) | Pace | Heavy Rail | Any Transit |
| --- | --- | --- | --- | --- |
| **not Available** | 124,548 | 115,880 | 116,128 | 73,724 |
|  | (80%) | (75%) | (75%) | (47%) |
| **Available** | 30,975 | 39,643 | 39,395 | 81,799 |
|  | (20%) | (25%) | (25%) | (53%) |
| **Sum:** | 155,523 | 155,523 | 155,523 | 155,523 |

Figure 2- Transit Availability for the Observed Trips

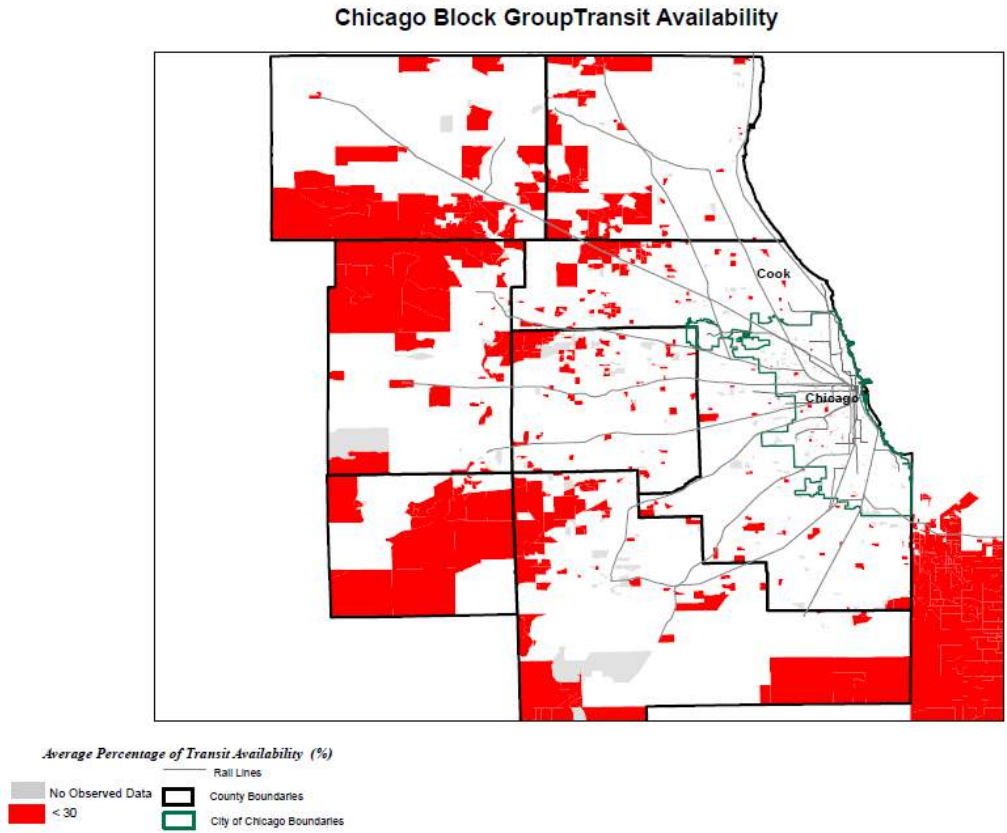

Traditionally when developing a mode choice model, travel time skims are generated from a regional travel demand model, and average travel times are used to represent zone to zone travel times. Some modelers have tried to improve this method by making the zones smaller. However as it can be seen in following graphs and tables while for many trips a transit alternative basically does not exist, assuming average travel times between zones is not the best estimation.

Table 4 indicates that almost 45,000 of the trips that, according to Google map, did not have any transit alternative were between zones that almost 27,000 other trips had at least one valid transit alternative (Table 4). This is another type of constraints that individuals face when choosing a mode for travel. That is to improve the accuracy of models, it is necessary to eliminate related alternatives when they do not exist in reality.

Moreover, Figure 3 also depicts the distribution of $\frac{STD}{Average}$ for zone to zone travel times (for zones with more than 5 observations) for different modes. As it can be seen the travel times between zones is highly a variable and follows a distribution. For example for the case of CTA, the average error is around 40% while for the case of auto (both for entire day and rush hour) the error is slightly higher. The distributions represented in this graph basically describe why assuming an average travel time skim that is generated from a regional travel demand model is not the best approach when generating alternative modes (The rush-hour distribution were plotted to find out whether variability of time throughout the day has been the root of dispersity).

Table 4- Number of trips with and without transit alternative

|  | CTA | Pace | Rail | Sum |
|---|---|---|---|---|
| **without** | 6,421 | 19,541 | 18,924 | 44,886 |
| **with** | 7,251 | 11,414 | 8,267 | 26,932 |

Figure 3 - Distribution of $\frac{standard\ deviation}{average}$ of zone to zone travel times for different modes of travel

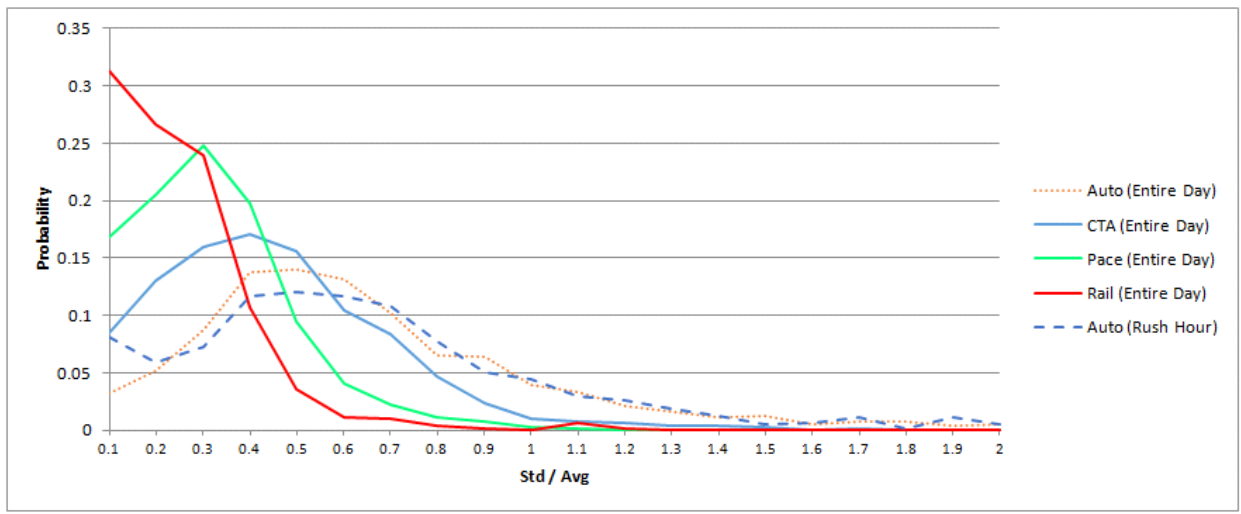

## Model Results

As discussed earlier, when it comes to selecting a mode by an individual, resource constraint issues including alternative mode availability have to play a significant role in preparing data for estimating mode choice models. Not only the availability of alternatives at spatial and temporal dimensions, but also, the accuracy of the actual alternatives for the observed trips and their attributes such as travel time need to be carefully considered in order to feed reliable data in to model estimation procedure. Furthermore, mode choice is not only dependent on a trip, but also the associated tour. For instance, if a tour of activities is engaged with one activity/trip being executed in the CBD area, this might affect the mode choice for the other trips in the tour. On the other hand, it is not acceptable to have one mode choice model for all the trips in a tour since subsequent trips in a tour have correlated mode decisions. For example, if an individual leaves home with his/her car would most likely use auto for the subsequent trips; therefore, it is logical to have multiple tour-based mode choice models for various trip types while considering the tour characteristics.

Considering the choice set formation and the alternative attribute generation approach used in this study, this section provides the output of an empirical multinomial logit (MNL) mode choice model estimated from the manipulated data. The model is only associated with the trip types that are home-based with home as their origin (The initial trip of the tours).

The results of the study suggests that valid choice set formation which is engaged with taking in to account individuals' resource constraints in selecting modes as well as accurate alternative attribute generation can significantly improve the fitness of the estimated models well beyond the impact of appropriate econometric formulations. Although MNL suffers from the IIA property and other econometric formulations could alleviate this limitation, the importance of sound alternative generation approach seems to make the estimated model surpass its counterparts.

Figure () displays the result of the estimated MNL model. 8 different modes are considered for this study; two non-motorized, two auto and four transit modes. The transit modes include the intracity and intercity available transit network in the greater Chicago area. A number of key variables were examined to specify the corresponding utility functions. Variables used to specify the utility function of walking mode include age, travel time and walking distance. For biking, due to the infrequency of data, travel time is the only determinant to form the utility function. Auto mode including drive and passenger could be dependent on a number of sociodemographic and economic conditions like gender, income, number of household members and number of vehicles as well as trip characteristics such as origin, destination, activity type, time of day, gas price and day of week. The data shows that people tend to use auto mode in case of longer travel times in spite of the gas fare increase which slows down the preference. Apparently, the data along with the model shows that female individuals drive more frequently than male ones. Furthermore, income is a significant factor in increasing the drive mode. Driving to the CBD area in the rush hours is a less frequent drive; therefore, if the destination is CBD, the parameter is negative. Activity type is another factor that affects the mode preference; shopping activities are more executed with the auto mode. On the other hand, trips for work related activities are less made with the passenger mode. Individuals are more inclined to drive on the weekends; less to use passenger mode. Passenger mode is more probable with higher number of household members and less with higher number of vehicles which is intuitive based on vehicle availability for driving instead of passenger mode. CTA mode which stands for the Chicago Transit Authority includes the bus and the subway network operating mainly in the city of Chicago. For CTA riders, the higher the number of transfers, income, and access/egress distance are, the less preference is remained for the ride; while increase in the travel time or the trip purpose being work-related makes the CTA more welcoming for the riders. PACE as an intercity transit network is less favored by higher income people. Also, long access, egress distances reduces the PACE ridership significantly. Finally, the heavy rail intercity transit network is divided in to two modes as slow and fast for non-motorized and motorized access respectively. Number of transfers, transit access distance and fare are the negative factors for the ridership; while intercity trips in the rush hour are frequently made with this heavy rail.

For the goodness of fit of the model, the likelihood ratio index was achieved based on the statistic suggested by McFadden (1974). This index although different in concept is analogous to the R-square in linear regression models. This model was estimated based on 80% of the survey sample (training data). The result shows that the likelihood ratio (0.73) is considerably higher than the typical mode choice models. This represents the significance of the choice set formation approach which is constructed by considering resource constraints not to mention the accuracy which is pursued in obtaining actual alternatives' attributes such as travel times, transit access distances and transit transfers.

$$R^2 = 1 - \frac{\ln L}{\ln L_0} \quad (1)$$

| Variable | Walk | Bike | Drive | Passenger | CTA | PACE | HRail Slow Access | HRail Fast Access |
|---|---|---|---|---|---|---|---|---|
| | | | **Home-based Mode Choice (Trip Origin is home)** | | | | | |
| Travel Time | -1.628 (-30.8) | -5.692 (-30.5) | 6.002 (44.56) | 6.002 (44.56) | 1.189 (26.97) | 1.189029 (26.97) | 1.189029 (26.97) | 1.189029 (26.97) |
| Number of Household members | | | | 0.949 (83.10) | | | | |
| Number of Household Vehicles | | | | -0.668 (-87.1) | | | | |
| Female (Y N) | | | 0.60318 (24.64) | | | | | |
| Number of Trip Transfers | | | | | -1.984761 (-21.2) | | -1.0269 (-3.38) | -2.78262 (-4.5) |
| Income (*10E-05) | | | 1.854156 (59.47) | | -0.194333 (-3.05) | -2.098 (-6.33) | | |
| Is any Trip of the Tour from City to the Suburbs or vice versa in the rush hour? (Y N) | | | | | | | 4.05 (15.12) | 4.05 (15.12) |
| Is any Destination in the tour CBD area in the rush hour? (Y N) | | | -1.639 (-13.94) | | | | | |
| Is the Trip for Shopping Purposes (Y N) | | | 1.2188 (26.03) | 0.542233 (10.69) | | | | |
| Is the Trip for Work Purposes (Y N) | | | | -1.93492 (-54.1) | 0.738 (12.69) | | | |
| Weekend (Y N) | | | 0.266 (5.92) | -0.091687 (-1.89) | | | | |
| Transit Access Distance | | | | | -0.330366 (-5.62) | -1.175 (-6.57) | -2.927 (-9.49) | -0.374 (-2.44) |
| Transit Egress Distance | | | | | -0.052364 (-0.95) | -1.532 (-9.49) | | |
| Destination Within Walking Distance (Y N) | 2.241 (78.5) | | | | | | | |
| Age Over 65 (Y N) | -0.923 (-19.67) | | | | | | | |
| Fare(Transit Fare or Gas Price) | | | -1.016 (-29.55) | -1.220 (-34.9) | | -0.848 (-11) | -1.694 (-21.58) | -1.694 (-21.58) |
| Log Likelihood | -52426 | | | | | | | |
| Observations | 88913 | | | | | | | |
| Likelihood Ratio Index | 0.736 | | | | | | | |

## Conclusion

This paper tried to demonstrate the importance of conserving resource constraints in microsimulation-based models by elaborating development of a mode choice model. A descriptive analysis was conducted on CMAP travel tracker survey to reveal the hidden constraints that individual's encounter which should be taken into account when developing a model or at simulation time. Google map website was also consulted to accurately generate alternative modes (if exist) and find their attributes. The resource constraints (availability of mode) and relatively accurate travel times and costs, along with other information, were then taken into account to generate a comprehensive choice set for model development. It was also emphasized that the mode choice should be developed within a tour-based framework. The result of the developed model shows the significant improvement in the likelihood ratio and the estimated parameters. For future work, combination of a more plausible mode choice modeling formulation such as nested logit along with the reliable data could enhance the quality of the modelling results even further. Plus, providing a mode choice model or a rule-based approach for trips other than the initial trip of the tours (starting from home) could complement the mode choice modeling, making it ready for application in an activity-based framework.

## References

1. Bhat C, Koppelman F. Activity-Based Modeling of Travel Demand. In: Hall R, editor. Handbook of Transportation Science. International Series in Operations Research & Management Science. 56: Springer US; 2003. p. 39-65.

2. Bradley M, Bowman JL, Griesenbeck B. SACSIM: An applied activity-based model system with fine-level spatial and temporal resolution. Journal of Choice Modelling. 2010;3(1):5-31.

3. Lemp J, McWethy L, Kockelman K. From Aggregate Methods to Microsimulation: Assessing Benefits of Microscopic Activity-Based Models of Travel Demand. Transportation Research Record: Journal of the Transportation Research Board. 2007;1994(-1):80-8.

4. Raney B, Cetin N, Völlmy A, Vrtic M, Axhausen K, Nagel K. An Agent-Based Microsimulation Model of Swiss Travel: First Results. Networks and Spatial Economics. 2003;3(1):23-41.

5. Recker W, Parimi A. Development of a Microscopic Activity-Based Framework for Analyzing the Potential Impacts of Transportation Control Measures on Vehicle Emissions. 1998.

6. Langerudi, Mehran Fasihozaman; Javanmardi, Mahmoud; Mohammadian, Abolfazl (Kouros); Sriraj, PS; "Choice Set Imputation", Transportation Research Record: Journal of the Transportation Research Board,2429,1,79-89,2014,Transportation Research Board of the National Academies,doi:10.3141/2429-09

7. Walker J. Making Household Microsimulation of Travel and Activities Accessible to Planners. Transportation Research Record: Journal of the Transportation Research Board. 2005;1931(-1):38-48.

8. Langerudi, Mehran Fasihozaman; Abolfazl, Mohammadian; Sriraj, PS; "Health and Transportation: Small Scale Area Association", Journal of Transport & Health, 2014, Elsevier, doi:10.1016/j.jth.2014.08.005

9.Malayath M, Verma A. Activity based travel demand models as a tool for evaluating sustainable transportation policies. Research in Transportation Economics. 2013;38(1):45-66.

10.Yagi S, Mohammadian A. An Activity-Based Microsimulation Model of Travel Demand in the Jakarta Metropolitan Area. Journal of Choice Modelling. 2010;3(1):32-57.

11. Salarian, Mehdi, and Hamid Hassanpour. "A new fast no search fractal image compression in DCT domain." Machine Vision, 2007. ICMV 2007. International Conference on. IEEE, 2007.

12.McNally MG. The Four Step Model. UC Irvine: Center for Activity Systems Analysis,; 2008.

13.McNally MG, Rindt CR. The Activity-Based Approach. UC Irvine: Center for Activity Systems Analysis,; 2007.

14. Fasihozaman Langerudi, Mehran; Hossein Rashidi, Taha; Mohammadian, Abolfazl; "Investigating the Transferability of Individual Trip Rates: Decision Tree Approach", Transportation Research Board 92nd Annual Meeting, 13-0218, 2013

15. Mehdi Salarian, "Accurate Localization in Dense Urban Area Using Google Street View Image", arXiv:1412.8496.

16.Auld J, Mohammadian A. Activity planning processes in the Agent-based Dynamic Activity Planning and Travel Scheduling (ADAPTS) model. Transportation Research Part A: Policy and Practice. 2012;46(8):1386-403.

17. Salarian, Mehdi, Ehsan Nadernejad, and H. M. Naimi. "A new modified fast fractal image compression algorithm." The Imaging Science Journal 61.2 (2013): 219-231.